\documentclass{article}
\usepackage{amssymb}
\usepackage{amsmath}
\usepackage{amsfonts}
\usepackage{graphicx}
\usepackage{epstopdf}
\usepackage{placeins}
\usepackage{color}
\usepackage{float}

\usepackage{tikz}
\usetikzlibrary{arrows.meta}
\usetikzlibrary{bending}

\begin{document}

\title{{\bf {On universal quantum dimensions of certain two-parameter series of representations.}\vspace{.2cm}}
\author{{\bf M.Y. Avetisyan} and {\bf R.L. Mkrtchyan}}
\date{ }
}

\maketitle





\begin{center}
  {\small {\it Yerevan Physics Institute, Yerevan, Armenia}}\\
\end{center}

\vspace{1cm}

\begin{abstract}
We  present the universal, in Vogel's sense, expression for the quantum dimension of Cartan product of an arbitrary number of adjoint and $X_2$ representations of simple Lie algebras. The same formula mysteriously gives quantum dimensions of some other representations of the same Lie algebra under permutations of universal parameters. We list these representations for exceptional algebras and stable versions  for classical algebras, when the rank of the classical algebra is sufficiently large w.r.t. the powers of representations. We show that  universal formulae can have singularities on Vogel's plane for some algebras, and that they give correct answers when restricted on appropriate lines on Vogel's plane. We note that the same irreducible representation can have several universal formulae for its (quantum) dimension, and discuss the implication of this phenomena on the Cohen - de Man method of calculation of universal formulae.

MSC classes: 17B20, 17B37, 57M25
\end{abstract}

\tableofcontents

\section{Introduction}

Present paper continues the efforts of a number of authors on revealing a "universality islands" in the theory of simple Lie algebras and their applications \cite{V0}-\cite{M16QD}. By "universality islands" we mean the ideas and formulae, which, particularly, represent different quantities in terms of so-called Vogel's universal parameters $\alpha, \beta, \gamma$, relevant up to scaling and permutations, correspondingly belonging to Vogel's plane \cite{LM1}. Simple Lie algebras correspond to specific points on that plane, given by Vogel's table, see Appendix A. These formulae connect objects (quantities) which belong to different simple Lie algebras, e.g. representations,(quantum) dimensions, volumes, etc. 

Theories of different simple Lie algebras are very similar, differing by the choice of the specific system of roots. Provided this choice is done, then construction of a large number of objects, such as the adjoint representation, i.e. the algebra itself, its finite-dimensional  representations, calculations of dimensions, etc., all is done by the same steps, with specific, for a given algebra, data used. However, no specific exact connection appears between objects in different simple algebras. For example, the adjoint representations of different simple Lie algebras are obviously "similar" objects, but e.g. their dimensions are for a first time given by the one, universal, (and analytic, in reasonable sense) formula in the paper of Vogel on his defined Universal Lie Algebra:

 \begin{eqnarray}
 \text{dim}\, \mathfrak{g} &=&-\frac{(2t-\alpha)(2t-\beta)(2t-\gamma)}{\alpha\beta\gamma} , t=\alpha+\beta+\gamma
 \end{eqnarray}

where universal Vogel parameters $\alpha, \beta, \gamma$ are given in the Vogel's table \ref{tab:V2}, see Appendix A for details. 

Actually some indication of universality appeared earlier from $N \leftrightarrow -N$ relation between sp(2n) and so(2n) algebras \cite{King} - \cite{MV11}. This relation became a part of the universality, however, in some aspects it is wider. Connected are representations of these two algebras with transposed Young diagrams, their dimensions are connected under the change of the sign of rank $N$ provided formulae for dimensions are represented as polynomials over $N$. This relation is part of the universality in the case of Young diagrams with even number of squares, however, it is valid for Young diagrams with odd number of squares also. There are also indications \cite{Cvit} on connection of spinor representations of $SO(2n)$ and metaplectic infinite-dimensional representation of $Sp(2n)$.

More details and examples of universal quantities are given in Appendix A. In the present paper we shall develop universality approach in the following directions. The main formulae are given in Section \ref{xkn}, particularly a universal expression for the quantum dimensions of Cartan (product of) powers of the adjoint and $X_2$ representations.The proof is presented in Appendix B. In Section \ref{Cas} we discuss some results on universal forms of eigenvalues of the Casimir operator. The possible application of the method of Cohen et al \cite{Cohen} for obtaining the universal expressions on the exceptional line to the full universal case is discussed in Conclusion.

Appendix A is devoted to some introductory information on Vogel's universality, Appendix B is devoted to the proof of the universal quantum dimension formula.

\section{Universal quantum dimension of Cartan product of powers of ad and $X_2$ representations} \label{xkn}

Consider the antisymmetric square of the adjoint representation. It is shown in \cite{V0} that it can be represented in a uniform form (i.e. for all simple Lie algebras) as

\begin{eqnarray}
\wedge^2 \mathfrak{g}=\mathfrak{g} \oplus X_2
\end{eqnarray}

The representation $X_2$ is irreducible w.r.t. the semidirect product of simple Lie algebra and the automorphism of the corresponding Dynkin diagram (see \cite{Cohen}) 
and it's highest weights in terms of fundamental ones are given in table \ref{hw}. We assume the numeration of nodes of the Dynkin diagram as in Mathematica program LieART, given in its description (\cite{FK}, page 11), with correction that the numeration of nods for $G_2$ is opposite to that given in \cite{FK}.

\FloatBarrier
\begin{table}[h] \label{hw}
	\caption{Highest weights of the adjoint ($\lambda_{ad}$) and $X_2$ ($\lambda_{X_2}$) representations in terms of fundamental weights for simple Lie algebras}
	\begin{tabular}{|c|c|c|}
		\hline
		&$\lambda_{ad}$&$\lambda_{X_2}$\\
		\hline
		$G_2$&$\omega_2$&$3\omega_1$\\
		\hline
		$F_4$&$\omega_1$&$\omega_2$\\
		\hline
		$E_6$&$\omega_6$&$\omega_3$\\
		\hline
		$E_7$&$\omega_1$&$\omega_2$\\
		\hline
		$E_8$&$\omega_7$&$\omega_6$\\
		\hline
		$A_1$&$2\omega$& 0 \\
		\hline
		$A_i, \,\, i>1$& $\omega_1+\omega_i$&$(2\omega_1+\omega_{i-1}) \oplus (\omega_2+2\omega_i)$ \\
		\hline
		$B_2$&$ 2\omega_2$&$\omega_1+2\omega_2$\\
		\hline
		$B_3$&$\omega_2$&$\omega_1+2\omega_3$\\
		\hline
		$B_i, \, i>3$&$\omega_2$&$\omega_1+\omega_3$\\
		\hline
		$C_i$&$2\omega_1$&$2\omega_1+\omega_2$\\
		\hline
		$D_4$&$\omega_2$&
		$\omega_1+\omega_3+\omega_4$\\
		\hline
		$D_i, \,\, i>4$&$\omega_2$&$\omega_1+\omega_3$\\
		\hline
	\end{tabular}
	\label{tab:hw}
\end{table}   
\FloatBarrier

Table \ref{tab:hw} needs a comment on the $A_i$ case. In that case $\lambda_{X_2}$ is not a highest weight, but a couple of highest weights, shown in the table. This is because representation $X_2$ is the sum of two irreducible representations of $A_i$, the highest weights of which are connected by automorphism of the Dynkin diagram. In that case the sum e.g. $\lambda_{X_2}+\lambda_{ad}$ should be understood as couple of weights, each is the sum of one vector from $\lambda_{X_2}$ and $\lambda_{ad}$.

Generally (see Proposition  below for the exact statement), the quantum dimension of irrep with highest weights $k\lambda_{X_2}+n\lambda_{ad}, \, k,n \in Z_+$ is given by the 
universal formula

\begin{multline}  \label{main}
   X(x,k,n,\alpha,\beta,\gamma)=\\
   L_{31}\cdot L_{32}\cdot L_{21s1}\cdot L_{21s2}\cdot L_{21s3}\cdot L_{10s1}\cdot
L_{10s2}\cdot L_{10s3}\cdot L_{11s1}\cdot L_{11s2}\cdot L_{11s3}\cdot L_{01}\cdot L_{c2}
\end{multline}

Multipliers $L_?$ look as follows  (see definition of sinh[x: in Appendix B)

$$
L_{31}=\sinh\left[\frac{x}{4}: \right. \frac{
-2 (\beta + \gamma)+\alpha(-4+3k+n)}{ 
2(2\alpha +\beta+\gamma)}$$

$$L_{32}=\sinh\left[\frac{x}{4}: \right. \frac{
-2 (\beta + \gamma)+\alpha(-3+3k+2n)}{ 
3 \alpha+2 (\beta+\gamma)}
$$

$$L_{21s1}=\sinh\left[\frac{x}{4}: \right.\prod _{i=1}^{2k+n} \frac{
-2(\beta+\gamma)+\alpha
(-5+i)}{-2\beta+
\alpha
(i-2)}
$$

$$L_{21s2}=\sinh\left[\frac{x}{4}: \right.\prod _{i=1}^{2k+n} \frac{
\beta+2\gamma-\alpha(-3+i)}
{\beta+\gamma -\alpha(i-2)}
$$

$$L_{21s3}=\sinh\left[\frac{x}{4}: \right. \frac{
2 \beta + \gamma+\alpha(3-2k-n)}{ 
3 \alpha+2 \beta+\gamma}
 $$

$$L_{10s1}=\sinh\left[\frac{x}{4}: \right.\prod _{i=1}^{k} \frac{  
   2\gamma-\alpha (i-3)}
{ -\alpha i}
$$

$$L_{10s2}=\sinh\left[\frac{x}{4}: \right.\prod _{i=1}^{ k} \frac{     
\beta+ \gamma-\alpha
(i-3)}
{ \beta-\alpha(i-2)}$$

$$L_{10s3}=\sinh\left[\frac{x}{4}: \right.\prod _{i=1}^{k} \frac{-2\beta+\alpha(i-3)}
{ \gamma-\alpha(i-2)}$$

$$L_{11s1}=\sinh\left[\frac{x}{4}: \right.\prod _{i=1}^{k+n} \frac{2\beta+\gamma-\alpha(i-4)}
{ \alpha(i+2)} $$

$$L_{11s2}=\sinh\left[\frac{x}{4}: \right.\prod _{i=1}^{k+n} \frac{\beta+\gamma-\alpha(i-2)}
{ \beta-\alpha (i-1)} $$

$$L_{11s3}=\sinh\left[\frac{x}{4}: \right.\prod _{i=1}^{k+n} \frac{-2\beta+\alpha(i-2)}
{ \gamma+\alpha(1-i)}$$

$$L_{01}=\sinh\left[\frac{x}{4}: \right. \frac{\alpha(1+n)}{ 
\alpha }$$

$$L_{c2}=\sinh\left[\frac{x}{4}: \right.\prod _{i=1}^k\frac{\gamma+2\beta-\alpha(i+k+n-4))}
{\alpha(i+k+n-2)-2\gamma}$$

{\bf Proposition.}

{\it The function  (\ref{main})	$X(x,k,n,\alpha,\beta,\gamma)$ on the points from Vogel's table is equal to the quantum dimensions of  representations of simple Lie algebras given in tables \ref{tab:xkna},  \ref{tab:xknae}.}

The proof of {\bf Proposition}
 is carried out case by case in Appendix B. I.e. for each set of parameters 
$\alpha, \beta, \gamma$ from Vogel's table 
the expression (\ref{main}) is compared with the Weyl formula for quantum
dimension (\ref{W}) for the corresponding algebra. The latter is the Weyl formula
 for characters, restricted to the Weyl line $x\rho$ (see e.g. \cite{DiF}, 13.170): 

\begin{eqnarray}\label{W}
D_Q^\lambda= 
\chi_{\lambda}(x\rho)= \prod_{\mu >0} \frac{\sinh(\frac{x}{2}(\mu,\lambda+\rho))}{\sinh(\frac{x}{2}(\mu,\rho))}
\end{eqnarray}

where $\lambda$ is the highest weight of the given irreducible representation, $\rho$ is the Weyl vector, the sum of the fundamental weights. Both sides of this formula are invariant w.r.t. the simultaneous rescaling (in "opposite directions") of the scalar product in algebra and the parameter $x$. The automorphism of the Dynkin diagram leads to the equality of quantum dimensions for representations with highest weights connected by the automorphism.

Note also  that the main formula (\ref{main}) is symmetric w.r.t. the switch of $\beta$ and $\gamma$ parameters, which can be established by careful inspection.

\FloatBarrier
\begin{table}[h]
	\caption{$X(x,k,n,\alpha,\beta,\gamma)$ for classical algebras}
	\begin{tabular}{|c|p{2cm}|p{2cm}|c|}
		\hline
		$k,n$&$0,n$&$1,n$ &$k,n \,\,(k>1)$ \\
		\hline
		$A_1$&$n \lambda_{ad}$&0&0\\  
		\hline
		$A_i,i\geq2$&$n\lambda_{ad}$&$\lambda_{X_2}+n\lambda_{ad} $&$k\lambda_{X_2}+n\lambda_{ad}$\\
		\hline
		$B_2$&$n\lambda_{ad}$&$\lambda_{X_2}+n\lambda_{ad}$&0\\  
		\hline
		$B_i, \, i>2$&$n\lambda_{ad}$&$\lambda_{X_2}+n\lambda_{ad} $&$k\lambda_{X_2}+n\lambda_{ad}$\\
		\hline
		$C_i, \, i>2$&$n\lambda_{ad}$&$\lambda_{X_2}+n\lambda_{ad} $&0\\
		\hline
		$D_i, \, i>3$&$n\lambda_{ad}$&$\lambda_{X_2}+n\lambda_{ad} $&$k\lambda_{X_2}+n\lambda_{ad}$\\
		\hline
	\end{tabular}
	\label{tab:xkna}
\end{table}     
\FloatBarrier

\FloatBarrier
\begin{table}[h]
	\caption{$X(x,k,n,\alpha,\beta,\gamma)$ for exceptional algebras}
	\begin{tabular}{|c|c|}
		\hline
		$k,n$&$k,n$\\
		\hline
		$L$&$k\lambda_{X_2}+n\lambda_{ad}$\\
		\hline
	\end{tabular}
	$L$ is any of exceptional simple Lie algebras.
	\label{tab:xknae}
\end{table}   

\FloatBarrier

\section{Permutation of the parameters}

According to the general features of the universality approach, the universal formulae make sense under permutations of Vogel parameters. In our case it means that permuting the parameters in the main formula (\ref{x2B}) we should obtain quantum dimensions of some other irreps of the same algebra. Indeed, it happens. The results are given in tables \ref{tab:xknbe} - \ref{tab:xkng}. In these tables all cases $(k,n)$, for which the corresponding function $X$ is non-zero, are listed - for any other combinations (except those known earlier, explicitly mentioned) corresponding function $X$ is zero. We propose these tables as conjectures, since we don't carry out the complete proof for all entries, only some random checks are done. 

The main peculiarity of these tables is that although in some cases the function $X$ is singular on a given point from Vogel's table, however, it always has a limit when restricted on the 
corresponding line. In table \ref{tab:xknbe} we show these cases explicitly, by symbol $E:$, but omit such an information in other tables. Note that $so(8)$ algebra belongs to both 
exceptional and orthogonal lines, and indeed both limits make sense, as shown below in the table \ref{tab:SO8}. As an example consider e.g. $X(x,1,2,4,-2,4)$, i.e.
the function with permuted parameters, corresponding to the $so(8)$ algebra ($X(x,1,2, \beta, \alpha, \gamma)$).
 Analyzing the $L_{?}$ terms for the corresponding set of the parameters, one finds, that the $X$ function has the following structure:
 
\begin{eqnarray}
X(x,1,2,\beta,\alpha,\gamma)=G(x) \sinh\left[x/4: \right. \,\, \frac{\alpha-2(\beta+\gamma)}{\alpha-\gamma} 
\end{eqnarray}

where $G(x)$ is regular at point  $(\alpha,\beta,\gamma)=(4,-2,4)$ in Vogel's plane. I.e. due to the $\frac{\alpha-2(\beta+\gamma)}{\alpha-\gamma}$ term the $X$ function is indeterminate at the $(4,-2,4)$ point. 
Let's move slightly our $so(8)$ point on Vogel's plane, taking  $\beta=4+p, \alpha=-2+q, \gamma=4$ with  $p,q \rightarrow 0$, thus the "indeterminate" term can be rewritten as
$$\frac{p-2q}{p}=1-2\frac{p}{q},$$ i.e. the value of $X$ at that point depends on the limit of $p/q$, which doesn't exist. However, if one approaches that point through $so$ line ($2\alpha+\beta=0$), then limit exists, since $p/q=-2$. When the same is done through $Exc$ line ($\gamma=2(\alpha+\beta)$), one have
$p/q=-1$. In both cases answer is reasonable and coincides with that for the  highest weight representations  presented in the Table 8, where also the permutations for $so(8)$ algebra in other possible
cases (not only for $(k,n)=(1,2)$) are analyzed.

\begin{table} 
	\caption{$X(x,k,n,\beta,\alpha,\gamma)$ for the exceptional algebras} \tiny
	\begin{tabular}{|c|c|c|c|c|c|}
		\hline
		
		$k,n$&$G_2$&$F_4$&$E_6$&$E_7$&$E_8$\\
		\hline
		1,0&$3\omega_1$&$\omega_2$ &  $\omega_3$ & $\omega_2$&$\omega_6$ \\
		\hline
		1,1& $\omega_1+\omega_2$&$\omega_3+\omega_4$&{\begin{tabular}{c}
				E:$(\omega_1+\omega_2)$\\
				$\oplus(\omega_4+\omega_5)$
			\end{tabular}} & $\omega_6+\omega_7$ & $\omega_8$ \\
		\hline
		1,2& 0 & $\omega_1+\omega_4$ & $\omega_3$&0 & $-\omega_8$ \\
		\hline
		1,3& 0 & 0 & 0& E:$-2\omega_6$ & $-\omega_6$ \\
		\hline
		1,4& 0 & 0 & -1 & 0 & 0  \\
		\hline
		1,5& 0 & 0 & 0 & 0 & 1 \\
		\hline
		2,0& 0 & $3\omega_4$ & E:$3\omega_1 \oplus 3\omega_5$ & 0 & 0 \\
		\hline
		2,1& 0 & 0 & $-\omega_3$ & E:$-\omega_6-\omega_7$ & 0 \\
		\hline
		2,2& 0 & 0 & $-\omega_6$ & $-\omega_5$ & $\omega_6$ \\
		\hline
		2,3& 0 & 0 & 0 & 0 & $\omega_7$ \\
		\hline
		3,0& 0 & 0 &  $-\omega_1-\omega_5$ & E:$-\omega_2$ & $\omega_8$ \\
		\hline
		3,1& 0 & 0 & 0 & E:$-\omega_1$ & $\omega_1$ \\
		\hline
		4,0& 0 & 0 & 0 & -1 & 0 \\
		\hline
	\end{tabular}
	\label{tab:xknbe} \normalsize
\end{table}

\begin{table} 
	\caption{$X(x,k,n,\gamma,\alpha,\beta)$ for the exceptional algebras} \scriptsize
	\begin{tabular}{|c|c|c|c|c|c|}
		\hline
		
		$k,n$&1,0&1,1&1,3&2,0&2,1\\
		\hline
		$G_2$&$3\omega_1$&$-3\omega_1$&1&$3\omega_1$&$\omega_2$\\
		\hline
		$F_4$&$\omega_2$ &$-\omega_2$&1&$\omega_2$&$\omega_1$\\
		\hline
		$E_6$& $\omega_3$&$-\omega_3$&1&$\omega_3$ &$\omega_6$  \\
		\hline
		$E_7$& $\omega_2$&$-\omega_2$&1&$\omega_2$&$\omega_1$ \\
		\hline
		$E_8$&$\omega_6$ &$-\omega_6$&1&$\omega_6$&$\omega_7$\\
		\hline
		
	\end{tabular}
	\label{tab:xknge} \normalsize
\end{table}

 \FloatBarrier
\begin{table} 
	\caption{$X(x,k,n,\beta,\alpha,\gamma)$ for the classical algebras. Data for $A_i, C_i$ are valid for sufficiently large rank $i$ (depending on $k,n$)} \scriptsize
	\begin{tabular}{|c|c|c|}
		\hline
		$k,n$&$1,n$&$k,n\,\, , \,\,k\geq 2$\\
		\hline
		$A_i$&$ (\omega_1+\omega_{1+n}+\omega_{i-1-n}) \oplus (\omega_{i}+\omega_{i-n}+\omega_{n+2})$ &$ (\omega_k+\omega_{k+n}+\omega_{i+1-2k-n})\oplus (\omega_{2k+n}+\omega_{i+1-k}+\omega_{i+1-k-n})$\\
		\hline
		$B_i$&$\omega_1+\omega_{2n+3}$&0 \\
		\hline
		$C_i$&$\omega_1+\omega_{n+1}+\omega_{n+2}$&$\omega_k+\omega_{k+n}+\omega_{2k+n}$\\
		\hline
		$D_i$&$\omega_1+\omega_{2n+3}$&0\\
		\hline
	\end{tabular}
	\label{tab:xknb} \normalsize
\end{table}
\FloatBarrier

In table \ref{tab:xknb} for the lower (i.e. not "sufficiently large") values of rank $i$ the function $X(x,k,n,\beta,\alpha,\gamma)$ still gives (quantum) dimensions of irreducible representations of corresponding algebra. However, the picture is chaotic and we don't try to clarify it. 

The case $(k,n)=(0,n)$ is considered in \cite{LM1}.

\FloatBarrier
\begin{table}[h]
	\caption{$X(x,k,n,\gamma,\alpha,\beta)$ for the classical algebras}
	\begin{tabular}{|c|c|}
		\hline
		$k,n>0$&$1,2$\\
		\hline
		$A_i$&-1\\
		\hline
		$L$&0\\
		\hline
	\end{tabular}
	$L$ is any classical algebra, except $A_i$
	\label{tab:xkng}
\end{table}   

\FloatBarrier

According to the table \ref{tab:xkng}, for the classical algebras,  $X(x,k,n,\gamma,\alpha,\beta)$  is non-zero  (besides previously known case $(k,n)=(k,0)$, \cite{AM}) for $(k,n)=(1,2), A_i$, only. In that case $X(x,k,n,\gamma,\alpha,\beta)=-1$.

\section{Permutation of the parameters for $so(8)$}

It is interesting to consider the case of $so(8)$ algebra separately. Its Dynkin diagram has the largest symmetry group, $S_3$, and belongs either to the orthogonal and the 
exceptional lines. So, one expects, that our formula for $X(x,k,n,\alpha,\beta,\gamma)$ should give reasonable results from both point of views, i.e. when restricted to either of these lines. Indeed it happens. In table \ref{tab:SO8} we present the results of the permutation and the restriction to each of the mentioned lines. 

\FloatBarrier

\begin{table} 
	\caption{Permutations of the parameters for $so(8)$ algebras} 
	\begin{tabular}{|c|c|c|c|c|c|c|c|c|}
		\hline
		
		Line&$k,n \, :$&1,0&1,1&1,2&1,3&2,0&2,1&3,0\\
		\hline
		Exc &$X(x,k,n,\beta,\alpha,\gamma)$&$\lambda_{X_2}$&$\lambda_{X_2}\oplus \lambda_{X_2}$&$\omega_1\oplus\omega_3\oplus\omega_4$&-2&$\lambda_{X_2}$&$\lambda_{ad}$&3\\
		\hline
		Exc & $X(x,k,n,\gamma,\alpha,\beta)$&$\lambda_{X_2}$ &$-\lambda_{X_2}$&0&1&$\lambda_{X_2}$&$\lambda_{ad}$&0\\
		\hline
		SO &$X(x,k,n,\beta,\alpha,\gamma)$& $\lambda_{X_2}$&$\lambda_{X_2}$&$\omega_{1}$&-1&0&0&0  \\
		\hline
		SO&$X(x,k,n,\gamma,\alpha,\beta)$& $\lambda_{X_2}$&0&0&0&0&0&0\\
				\hline		
	\end{tabular}
	\label{tab:SO8} 
\end{table}  

\FloatBarrier

All these results are in agreement with the results of paper \cite{AM} for $(k,n)=(k,0)$ case, and with the $N\leftrightarrow -N$ duality for $SO(N)/Sp(N)$ algebras \cite{King,Cvitbook,Mkr,MV11}.

\section{The universal eigenvalues of Casimir operator}\label{Cas}

Among interesting quantities, which can have universal form, are the eigenvalues of Casimir operator. See, e.g. \cite{V0,LM1,Cohen} for universal eigenvalues of second and fourth Casimir operators on adjoint, $Y_k(,)$ and other representations. In \cite{MSV} the generating function of higher Casimir operators on the adjoint representation is calculated in 
the universal form. Here we would like to present the results of \cite{A19} for representations, considered in previous sections, namely the universal form for Casimir operator eigenvalues on the representation with quantum dimension given by $X(k,n,\alpha,\beta,\gamma)$ is:

\begin{eqnarray}
C_{k,n}=\alpha(3k-3k^2+n-n^2-3kn)+t(4k+2n)
\end{eqnarray}

This coincides with eigenvalues, given in \cite{Cohen} for representations (Cartan products of) $X_2,X_2^2,gX_2,g^2X_2$ (with Casimir's eigenvalues $C_{1,0}, C_{2,0}, C_{1,1}, C_{1,2}$, respectively), which correspond to the representations $X_2, H, C, G$ (in notations of \cite{Cohen}) respectively. Further check of this formula, e.g. comparison with expressions from \cite{MV11} are presented in \cite{A19}. Particularly, permutations of parameters give eigenvalues, which coincide with eigenvalues of permuted representations from previous Section.

\section{Conclusion}

There is no a regular method of obtaining universal formulae. One possibility is the method of Cohen and de Man \cite{Cohen} of obtaining the universal formulae for dimensions of representations, appearing in the decomposition of a given power of adjoint representation. They developed and applied this method for obtaining dimensional formulae, "universal" on the exceptional line. That is, they depend on parameter on that line, e.g. as if they were obtained from universal formula with substitution (e.g.) $\gamma=2, \alpha=x, \beta=1-x$. Then one considers all representations, appearing in the powers of the adjoint (up to some fixed power), their products and decompositions of that products. These decompositions translate into relations between dimensions, or quantum dimensions of that representations. With some additional relations, including Casimir operator's eigenvalues, one can hope to obtain sufficient number of relations, to find all dimensions in the universal (on exceptional line) form, since the dimensions of the first few (three, actually) powers of the adjoint are found already in the universal form. We would like to note that this method doesn't work in true universal case, i.e. on the entire Vogel's plane. The point is that there exist representations, for a given simple Lie algebra, with more than one universal formula for their dimensions. 

Consider some representation of a given simple Lie algebra and assume that it has a universal representation for its dimension. As an example consider the adjoint representation of $sl(n)$ algebra. It has a universal representation for its dimension $n^2-1$:

\begin{eqnarray}
-\frac{(2t-\alpha)(2t-\beta)(2t-\gamma)}{\alpha\beta\gamma}
\end{eqnarray}

Among representations, appearing in the decomposition of the square of the adjoint, there is representation $Y_2(\gamma)$ with a universal formula for its dimension: 

\begin{eqnarray}
 \text{dim} \, Y_2(\alpha)&=& \frac{\left(  2t  - 3\alpha \right) \,\left( \beta - 2t \right) \,\left( \gamma - 2t \right) \,t\,\left( \beta + t \right) \,
      \left( \gamma + t \right) }{\alpha^2\,\left( \alpha - \beta \right) \,\beta\,\left( \alpha - \gamma \right) \,\gamma}
\end{eqnarray}

It is easy to check that in the case of $sl(n)$ algebra the dimension of $Y_2(\gamma)$ is the same, $n^2-1$, and they are actually the same representations, since there is no other representation with that dimension. Similarly, one can have other examples of a representation, for a given algebra, which has two universal origins. Evidently, this phenomena prevents one from using Cohen - de Man method, since for such representations it is not clear which formula exactly should be used. Further examples of this phenomena can be considered, when one has different symmetry groups acting on the "same" representation with different universal origins. Example of different symmetry groups is two $S_3$ groups, discovered in \cite{V0}.  

We assume all this to be the consequence of the existence of zero divisors (discovered in \cite{V}) in Vogel's $\Lambda$-algebra, which is the ring of the coefficients in his Universal Lie Algebra.The existence of such divisors leads to the impossibility of decomposition into irreducible modules, in the universal form. Of course, this fact doesn't prevent the existence of specific universal formulae. 

We assume that the regular method of derivation of universal formulae can be obtained by deeper understanding of the structure of Vogel's $\Lambda$-algebra and generalization of Vogel's approach based on Universal Lie Algebra. Particularly, the rules of restriction on $SO/Sp/Exc$ lines, in tables above, should be {\bf derived} within this (or any other) approach.

\section{Acknowledgments}

The work of MA was fulfilled within the Regional Doctoral Program on Theoretical and Experimental Particle Physics Program 
sponsored by VolkswagenStiftung.
The work of MA and RM is partially supported by the Science Committee of the Ministry of Science 
and Education of the Republic of Armenia under contract  18T-1C229. We are grateful to organizers of workshop "Supersymmetries and Quantum Symmetries – SQS'19" (Yerevan), where these results are reported, for invitation.

\section{Appendix A. Vogel's parameters}

Vogel's Universal Lie Algebra among other things leads to a parameterization of simple Lie algebras, given in the following table \ref{tab:V2}

\FloatBarrier
\begin{table}[ht] 
	\caption{Vogel's parameters for simple Lie algebras: lines}
	\begin{tabular}{|r|r|r|r|r|r|} 
		\hline Algebra/Parameters & $\alpha$ &$\beta$  &$\gamma$  & $t$ & Line \\ 
		\hline  $\mathfrak {sl}_{N}$  & -2 & 2 & $N$ & $N$ & $\alpha+\beta=0$ \\ 
		\hline $\mathfrak {so}_{N}$ & -2  & 4 & $N-4$ & $N-2$ & $ 2\alpha+\beta=0$ \\ 
		\hline  $ \mathfrak {sp}_{N}$ & -2  & 1 & $N/2+2$ & $N/2+1$ & $ \alpha +2\beta=0$ \\ 
		\hline $Exc(n)$ & $-2$ & $2n+4$  & $n+4$ & $3n+6$ & $\gamma=2(\alpha+\beta)$\\ 
		\hline 
	\end{tabular}
	
	{For the exceptional 
		line $n=-2/3,0,1,2,4,8$ for $\mathfrak {g}_{2}, \mathfrak {so}_{8}, \mathfrak{f}_{4}, \mathfrak{e}_{6}, \mathfrak {e}_{7},\mathfrak {e}_{8} $, 
		respectively.} \label{tab:V2}
\end{table} 
\FloatBarrier

The simplest origin of Vogel's table is the universal decomposition of the symmetric square of the adjoint representation \cite{V0}:

\begin{eqnarray}\label{sad}
S^2 \mathfrak{g}=1 \oplus Y_2(\alpha) \oplus Y_2(\beta) \oplus Y_2(\gamma)
\end{eqnarray}

 Denote $2t$ as the value of the second Casimir operator on the adjoint representation $\mathfrak{g}$. Then parameterize the values of the same operator on representations in (\ref{sad}) as $4t-2\alpha, 4t-2\beta, 4t-2\gamma$, that will be the definition of three Vogel's parameters.

A number of universal formulae were derived. Dimension formulae are given in \cite{V0,V,LM1}. Universal formulae were derived for compact groups volume \cite{M13,KM} and for the anomaly of Vogel's permutation symmetry \cite{KM}. Universal formulae were discovered in applications, mainly in Chern-Simons theory: partition function \cite{MV,MV11}, knot polynomials  (Wilson loops) \cite{MMM}, central charge, etc.

\section{Appendix B. The proof of  Proposition }

\subsection{Notation sinh[x:} 
We  use the following notation:

\begin{eqnarray}
a\sinh\left[x: \right. \,\, \frac{A\cdot B...}{M\cdot N...}\equiv a\frac{\sinh(xA)\sinh(xB)...}{\sinh(xM)\sinh(xN)...}
\end{eqnarray}
where $x,a,A,B,...,M,N,...$ are numbers (dots between  are not necessary, provided no ambiguity arises). For example

\begin{eqnarray}
2\sinh\left[\frac{x}{4}: \right. \,\, \frac{1\cdot 4}{2}= 2\frac{\sinh(\frac{x}{4})\sinh(\frac{4x}{4})}{\sinh(\frac{2x}{4})}
\end{eqnarray}

One can derive simple rules which this notation obeys. E.g. 

\begin{eqnarray}
\left( \sinh\left[x: \right. \,\, A\cdot B \right) \left( \sinh \left[x: \right. M\cdot N\right) = \sinh\left[x: \right. \,\, A\cdot B \cdot M\cdot N
\end{eqnarray}

\subsection{Proof of Proposition}

The proof is carried out case by case:  
for each set of parameters 
$\alpha, \beta, \gamma$ from Vogel's table  (except  $C_n$) we
compare the expression (\ref{main})  with the  quantum
dimension obtained by Weyl formula (\ref{W}) for the corresponding algebra.

\subsection{The $A_N$ Case}
 Substituting $\alpha=-2,\beta=2,\gamma=N+1$
in the 
$L$-terms, one gets
$$L_{31}=\sinh\left[\frac{x}{2}: \right.
\frac{  
3k+n+N-2}{ N-2 },$$

$$L_{32}=\sinh\left[\frac{x}{2}: \right.
\frac{ N+2n+3k-1}{N-1},$$

$$L_{21s1}=\sinh\left[\frac{x}{2}: \right.
\frac{  
  (N-2) \cdot  
\left(N-1 \right)\dots  
\left(N+n+2k-3  \right)}{1\cdot2\dots (2k+n)},$$

$$L_{21s2}=\sinh\left[\frac{x}{2}: \right.
\frac{  
\left(N-1\right)\cdot  
  N \dots  
\left(N+2k+n-2\right)}{ N/2\cdot (N/2+1) \dots (N/2+2k+n-1)},$$

$$L_{21s3}=\sinh\left[\frac{x}{2}: \right.
\frac{ N/2+2k+n-1}{N/2-1},$$

$$L_{10s1}=\cdot \sinh\left[\frac{x}{2}: \right.
\frac{ (N-2)\cdot
   (N-1)\cdot N\dots 
(N+k-3)  }{ 1
\cdot 2 \dots
k},$$

$$L_{10s2}=\cdot \sinh\left[\frac{x}{2}: \right.
\frac{ (N/2-1)\cdot
   N/2\cdot (N/2+1)\dots 
(N/2+k-2)  }
{(\alpha+\beta)
\cdot 1 \dots
k},$$

$$L_{10s3}=\cdot \sinh\left[\frac{x}{2}: \right.
\frac{ -2(\alpha+\beta)\cdot
   1\cdot 2\dots 
k }{ (N/2-1)
\cdot N/2 \dots
(N/2+k-2)},$$

$$L_{11s1}=\sinh\left[\frac{x}{2}: \right.
\frac{  
\left(N/2-1\right)\cdot  
\left(N/2\right)\dots  
\left(N/2+k+n-2 \right)}
{ 2\cdot3\dots (k+n+1)},$$

$$L_{11s2}=\sinh\left[\frac{x}{2}: \right.
\frac{ N/2 \cdot
\left(N/2+1\right) \dots  
\left(N/2+k+n-1 \right)}
{ 1\cdot 2\cdot 3\dots (k+n)},$$

$$L_{11s3}=1/L_{11s2}$$

$$L_{01}=\sinh\left[\frac{x}{2}: \right. 
\frac{1+n}{1}$$

$$L_{c2}=\sinh\left[\frac{x}{2}: \right.
\frac{ (N/2+k+n-1) \cdot
\left(N/2+k+n)\right) \dots  
\left(N/2+2k+n-2 \right)}
{ (N+k+n-1)\cdot (N+k+n)\dots (N+2k+n-2)},$$

The product of all these terms gives 

\begin{multline*}
   X(x,k,n,-2,2,N+1)=\\
   =L_{31}\cdot L_{32}\cdot L_{21s1}\cdot L_{21s2}\cdot L_{21s3}\cdot L_{10s1}\cdot
L_{10s2}\cdot L_{10s3}\cdot L_{11s1}\cdot L_{11s2}\cdot L_{11s3}\cdot L_{01}\cdot L_{c2}=\\
2\cdot \sinh\left[\frac{x}{2}: \right.
\frac{(N-1)^2 N^3(N+1)^3\dots(N+k-2)^3(N+k-1)^2\dots (N+k-1+n)^2}
 {1^3\cdot2^3\dots k^3}\cdot\\
 \frac{(N+k+n)\dots (N+2k+n-2)}
 {(k+1)^2\cdot(k+2)^2\dots(k+n)^2(k+n+1)^2(k+n+2)\dots (2k+n)}
\end{multline*}

which equals to the double of the expression of the Weyl formula, written for $\lambda=(2k+n)\omega_1+k\omega_{N-1}+n\omega_N$ highest weight representation of $A_N$ algebra, as expected.

\subsection{The $B_N$ Case}

 For this case we should substitute $\alpha=-2,\beta=4,\gamma=2N-3$, so 
 
$$L_{31}=\sinh\left[\frac{x}{2}: \right.
\frac{2N+3k+n-3}
{2N-3},$$

$$L_{32}=\sinh\left[\frac{x}{2}: \right.
\frac{2N+3k+2n-2}
{2N-2},$$

$$L_{21s1}=\sinh\left[\frac{x}{2}: \right.
\frac{  
  (2N-3) \cdot  
\left(2N-2 \right)\dots  
\left(2N+n+2k-4  \right)}{3\cdot4\dots (2+2k+n)},$$

$$L_{21s2}=\sinh\left[\frac{x}{2}: \right.
\frac{  
\left(2N-3\right)\cdot  
  (2N-2)\dots  
\left(2N+2k+n-4\right)}{ (N-1/2)\cdot (N+1/2) \dots (N+2k+n-3/2)},$$

$$L_{21s3}=\sinh\left[\frac{x}{2}: \right.
\frac{ N+2k-1/2}{N-1/2},$$

$$L_{10s1}=\cdot \sinh\left[\frac{x}{2}: \right.
\frac{ (2N-5)\cdot
   (2N-4)\dots 
(2N+k-6)  }{ 1
\cdot 2 \dots
k},$$

$$L_{10s2}=\cdot \sinh\left[\frac{x}{2}: \right.
\frac{ (N-3/2)\cdot
   (N-1/2)\dots 
(N+k-5/2)  }
{1\cdot 2 \dots
k},$$

$$L_{10s3}=\cdot \sinh\left[\frac{x}{2}: \right.
\frac{ 2\cdot
   3\cdot 4\dots 
(k+1) }{ (N-5/2)
\cdot (N-3/2) \dots
(N+k-7/2)},$$

$$L_{11s1}=\sinh\left[\frac{x}{2}: \right.
\frac{  
\left(N-1/2\right)\cdot  
\left(N+1/2\right)\dots  
\left(N+k+n-3/2 \right)}
{ 2\cdot3\dots (k+n+1)},$$

$$L_{11s2}=\sinh\left[\frac{x}{2}: \right.
\frac{ (N-1/2) \cdot
\left(N+1/2\right) \dots  
\left(N+k+n-3/2 \right)}
{ 2\cdot 3\cdot 4\dots (k+n+1)},$$

$$L_{11s3}=\sinh\left[\frac{x}{2}: \right.
\frac{ 3 \cdot
4 \dots  (k+n+2)}
{ (N-3/2)\cdot (N-1/2)\dots (N+k+n-5/2)},$$

$$L_{01}=\sinh\left[\frac{x}{2}: \right. 
\frac{1+n}{1}$$

$$L_{c2}=\sinh\left[\frac{x}{2}: \right.
\frac{ (N+k+n-1/2) \cdot
\left(N+k+n+1/2)\right) \dots  
\left(N+2k+n-3/2 \right)}
{ (2N+k+n-4)\cdot (2N+k+n-3)\dots (2N+2k+n-5)},$$

So, the product of all  $L$-terms is:
\begin{multline}\label{x2B}
     X(x,k,n,-2,4,2N-3)=\\
     L_{31}\cdot L_{32}\cdot L_{21s1}\cdot L_{21s2}\cdot L_{21s3}\cdot L_{10s1}\cdot
L_{10s2}\cdot L_{10s3}\cdot L_{11s1}\cdot L_{11s2}\cdot L_{11s3}\cdot L_{01}\cdot L_{c2}=\\
\sinh\left[\frac{x}{2}: \right.
\frac{(2N-5)(2N-4)^2(2N-3)^3\dots(2N+k-6)^3(2N+k-5)^2\dots (2N+k+n-5)^2}
 {1^3\cdot2^3\dots k^3 (k+1)^2\cdot(k+2)^2\dots(k+n)^2(k+n+1)^2 }\cdot\\
 \frac{(2N+k+n-4)\dots(2N+2k+n-5)(2N+2k+n-4)^2(k+n)(n+1)(k+n+2)(k+N-5/2)}
 {(k+n+2)\dots (2k+n+2)(N-1/2)(N-3/2)(N-5/2)}\cdot\\
  \frac{(k+n+N-3/2)(2k+n+N-1/2)(3k+n+2N-3)(3k+2n+2N-2)}
 {(2N-3)(2N-4)(2N-2)}
 \end{multline}

It coincides with the Weyl formula, written for $\lambda=k\omega_1+n\omega_2+k\omega_3$ highest weight representation of $B_N$ algebra.

\subsection{The $C_N$ Case}

 The Vogel parameters in this case are $\alpha=-2,\beta=1,\gamma=N+2$, and we notice, that for $k\geq2$ and for any $n$, the formula gives $0$, due to the contribution of $L_{10s3}$ term.
 So, we observe the $L_{?}$ terms for $k=1$ and any $n$.
 Thus, one has
 
$$L_{31}=\sinh\left[\frac{x}{2}: \right.
\frac{2N+4n+6}
{2N},$$

$$L_{32}=\sinh\left[\frac{x}{2}: \right.
\frac{2N+3k+2n-2}
{2N-2},$$

$$L_{21s1}=\sinh\left[\frac{x}{2}: \right.
\frac{  
  (N-1) \cdot  
N \dots  
\left(N+n \right)}{(-\alpha-2\beta)/2\cdot1\dots (n+1)},$$

$$L_{21s2}=\sinh\left[\frac{x}{2}: \right.
\frac{  
\left(N+1/2\right)\cdot  
  (N+3/2)\dots  
\left(N+n+3/2\right)}{ (N/2+1/2)\cdot (N/2+3/2) \dots (N/2+n+3/2)},$$

$$L_{21s3}=\sinh\left[\frac{x}{2}: \right.
\frac{ N/2+n+1}{N/2-1},$$

$$L_{10s1}=\cdot \sinh\left[\frac{x}{2}: \right.
\frac{ N  }{1},$$

$$L_{10s2}=\cdot \sinh\left[\frac{x}{2}: \right.
\frac{ N/2-1/2 }
{1/2},$$

$$L_{10s3}=\cdot \sinh\left[\frac{x}{2}: \right.
\frac{ 1 }{ N/2},$$

$$L_{11s1}=\sinh\left[\frac{x}{2}: \right.
\frac{  
\left(N/2-1\right)\cdot  
N/2\dots  
\left(N/2+n-1 \right)}
{ 2\cdot3\dots (n+2)},$$

$$L_{11s2}=\sinh\left[\frac{x}{2}: \right.
\frac{ (N/2+1/2) \cdot
\left(N/2+3/2\right) \dots  
\left(N/2+n+1/2 \right)}
{1/2\cdot 3/2\cdot 4/2\dots (n+1/2)},$$

$$L_{11s3}=\sinh\left[\frac{x}{2}: \right.
\frac{ (-\alpha-2\beta)/2 \cdot
1 \dots  n}
{ (N/2+1)\cdot (N/2+2)\dots (N/2+n+1)},$$

$$L_{01}=\sinh\left[\frac{x}{2}: \right. 
\frac{1+n}{1}$$

$$L_{c2}=\sinh\left[\frac{x}{2}: \right.
\frac{ N/2+n}
{ N+n+2},$$

And the product is
\begin{multline}\label{b}
     X(x,k,n,-2,4,2N-3)=\\
     L_{31}\cdot L_{32}\cdot L_{21s1}\cdot L_{21s2}\cdot L_{21s3}\cdot L_{10s1}\cdot
L_{10s2}\cdot L_{10s3}\cdot L_{11s1}\cdot L_{11s2}\cdot L_{11s3}\cdot L_{01}\cdot L_{c2}=\\
\sinh\left[\frac{x}{2}: \right.
\frac{2N(2N+1)(2N+2)\dots(2N+2n+1)(N-1)(2n+3)(2N+2n+3)(2N+4n+6)}
 {1^2\cdot2\cdot3\dots(2n+4)(N+2n+3)}
 \end{multline}

Which coincides with the Weyl formula, written for $\lambda=(2+2n)\omega_1+\omega_2$ highest weight representation of $C_N$ algebra.

\subsection{The $D_N$ Case}
 For this case
we substitute $\alpha=-2,\beta=4,\gamma=2N-4$, so $L$-terms become

$$L_{31}=\sinh\left[\frac{x}{2}: \right.
\frac{2N+3k+n-4}
{2N-4},$$

$$L_{32}=\sinh\left[\frac{x}{2}: \right.
\frac{2N+3k+2n-3}
{2N-3},$$

$$L_{21s1}=\sinh\left[\frac{x}{2}: \right.
\frac{  
  (2N-4) \cdot  
\left(2N-3 \right)\dots  
\left(2N+n+2k-5  \right)}{3\cdot4\dots (2+2k+n)},$$

$$L_{21s2}=\sinh\left[\frac{x}{2}: \right.
\frac{  
\left(2N-4\right)\cdot  
  (2N-3)\dots  
\left(2N+2k+n-5\right)}{ (N-1)\cdot N \dots (N-2+2k+n)},$$

$$L_{21s3}=\sinh\left[\frac{x}{2}: \right.
\frac{ N+2k+n-1}{N-1},$$

$$L_{10s1}=\cdot \sinh\left[\frac{x}{2}: \right.
\frac{ (2N-6)\cdot
   (2N-5)\dots 
(2N+k-7)  }{ 1
\cdot 2 \dots
k},$$

$$L_{10s2}=\cdot \sinh\left[\frac{x}{2}: \right.
\frac{ (N-2)\cdot
   (N-1)\dots 
(N+k-3)  }
{1\cdot 2 \dots
k},$$

$$L_{10s3}=\cdot \sinh\left[\frac{x}{2}: \right.
\frac{ 2\cdot
   3\cdot 4\dots 
(k+1) }{ (N-3)
\cdot (N-2) \dots
(N+k-4)},$$

$$L_{11s1}=\sinh\left[\frac{x}{2}: \right.
\frac{  
\left(N-1\right)\cdot  
N\dots  
\left(N+k+n-2 \right)}
{ 2\cdot3\dots (k+n+1)},$$

$$L_{11s2}=\sinh\left[\frac{x}{2}: \right.
\frac{ (N-1) \cdot
N \dots  
\left(N+k+n-2 \right)}
{ 2\cdot 3\cdot 4\dots (k+n+1)},$$

$$L_{11s3}=\sinh\left[\frac{x}{2}: \right.
\frac{ 3 \cdot
4 \dots  (k+n+2)}
{ (N-2)\cdot (N-1)\dots (N+k+n-3)},$$

$$L_{01}=\sinh\left[\frac{x}{2}: \right. 
\frac{1+n}{1}$$

$$L_{c2}=\sinh\left[\frac{x}{2}: \right.
\frac{ (N+k+n-1) \cdot
\left(N+k+n)\right) \dots  
\left(N+2k+n-2 \right)}
{ (2N+k+n-5)\cdot (2N+k+n-4)\dots (2N+2k+n-6)},$$

Overall, for $X(x,k,n,-2,4,2N-4)$ one gets
\begin{multline}\label{x2D}
     X(x,k,n,-2,4,2N-4)=\\
     L_{31}\cdot L_{32}\cdot L_{21s1}\cdot L_{21s2}\cdot L_{21s3}\cdot L_{10s1}\cdot
L_{10s2}\cdot L_{10s3}\cdot L_{11s1}\cdot L_{11s2}\cdot L_{11s3}\cdot L_{01}\cdot L_{c2}=\\
\sinh\left[\frac{x}{2}: \right.
\frac{(2N-6)(2N-5)(2N-4)^2(2N-3)^2(2N-2)^3\dots(2N+k-7)^3}
 {1^3\cdot2^3\dots k^3 (k+1)^2 }\times\\
\frac{(2N+k-6)^2\dots (2N+k+n-6)^2}
 {(k+2)^2\dots(k+n+1)^2 }\times\\
  \frac{(2N+k+n-5)\dots(2N+2k+n-6)(2N+2k+n-5)^2(N+2k+n-1)(n+1)(k+n+2)}
 {(k+n+2)(2k+n+2)}\times\\
  \frac{(k+1)(N+k-3)(N+k+n-2)}
 {(N-1)(N-2)(N-3)}
 \end{multline}

This coincides with the Weyl formula answer for $\lambda=k\omega_1+n\omega_2+k\omega_3$ highest weight representation.

\subsection{$G_2$}
For $G_2$ exceptional algebra Vogel's parameters 
take  values $\alpha=-2, \beta=10/3, \gamma=8/3$.
Substituting them in the $L$-terms, one has

$$L_{31}=\sinh\left[\frac{x}{2}: \right.
\frac{3k+n+2}{2},$$

$$L_{32}=\sinh\left[\frac{x}{2}: \right.\frac{3k+2n+3}{3}
,$$
$$L_{21s1}\times L_{21s2}=1,$$

$$L_{21s3}=\sinh\left[\frac{x}{2}: \right.\frac{5/3+2k+n}{5/3}
,$$

$$L_{10s1} \times L_{10s2}=1,$$

$$L_{10s3}=\sinh\left[\frac{x}{2}: \right.\frac{4/3\cdot7/3
\cdot10/3\dots(k+1/3)}
{1/3\cdot 4/3\dots\ (k-2/3)}=\sinh\left[\frac{x}{2}: \right.\frac{k+1/3}
{1/3},$$

$$L_{11s1} \times L_{11s2}=1,$$

$$L_{11s3}=\sinh\left[\frac{x}{2}: \right.\frac{7/3\cdot10/3
\dots(k+n+4/3)}
{4/3\cdot 7/3\dots\ (k+n+1/3)}=\sinh\left[\frac{x}{2}: \right.\frac{k+n+4/3}
{4/3},$$

$$L_{01}=\sinh\left[\frac{x}{2}: \right. 
\frac{1+n}{1}$$

$$L_{c2}=1$$

\begin{multline*}
 X(x,k,n,-2,10/3,8/3)=\\
 \sinh\left[\frac{x}{2}: \right.
  \frac{(3k+n+2)(3k+2n+3)(k+1/3)(k+n+4/3)(2k+n+5/3)(n+1)}
{1\cdot 1/3\cdot 4/3 \cdot 2 \cdot 3 \cdot 5/3},
\end{multline*}

which coincides with the expression the Weyl 
formula (\ref{W}) gives for quantum dimension of $G_2$ algebra.

\subsection{$F_4$}
In this case we have $\alpha=-2, \beta=5, \gamma=6$

$$L_{31}=\sinh\left[\frac{x}{2}: \right.
\frac{3k+n+7}{7},$$

$$L_{32}=\sinh\left[\frac{x}{2}: \right.\frac{3k+2n+8}{8}
,$$

$$L_{21s1}=\sinh\left[\frac{x}{2}: \right.\frac{7\cdot8
\cdot9\dots(6+2k+n)}
{4\cdot 5\dots\ (3+2k+n)}=\sinh\left[\frac{x}{2}: \right.\frac{(4+2k+n)(5+2k+n)(6+2k+n)}
{4\cdot5\cdot6},$$

$$L_{21s2}=\sinh\left[\frac{x}{2}: \right.\frac{(9/2+2k+n)(11/2+2k+n)}{9/2\cdot 11/2}
,$$

$$L_{21s3}=\sinh\left[\frac{x}{2}: \right.\frac{5+2k+n}{5}
,$$

$$L_{10s1}=\sinh\left[\frac{x}{2}: \right.\frac{4\cdot5
\cdot6\dots(3+k)}
{1\cdot 2\dots\ k}=\sinh\left[\frac{x}{2}: \right.\frac{(1+k)(2+k)(3+k)}
{1\cdot2\cdot3},$$

$$L_{10s2}=\sinh\left[\frac{x}{2}: \right.\frac{7/2\cdot9/2
\cdot10/2\dots(5/2+k)}
{3/2\cdot 5/2\dots\ (1/2+k)}=\sinh\left[\frac{x}{2}: \right.\frac{(3/2+k)(5/2+k)}
{3/2\cdot5/2},$$

$$L_{10s3}=\sinh\left[\frac{x}{2}: \right.\frac{3\cdot4
\cdot5\dots(k+2)}
{2\cdot 3\dots\ (k+1)}=\sinh\left[\frac{x}{2}: \right.\frac{k+2}
{2},$$

$$L_{11s1}=\sinh\left[\frac{x}{2}: \right.\frac{5\cdot6
\cdot7\dots(4+k+n)}
{2\cdot 3\dots\ (k+n+1)}=\sinh\left[\frac{x}{2}: \right.\frac{(k+n+2)(k+n+3)(k+n+4)}
{2\cdot3\cdot4},$$

$$L_{11s2}=\sinh\left[\frac{x}{2}: \right.\frac{9/2\cdot11/2
\cdot13/2\dots(7/2+k+n)}
{5/2\cdot 7/2\dots\ (3/2+k+n)}=\sinh\left[\frac{x}{2}: \right.\frac{(5/2+k+n)(7/2+k+n)}
{5/2\cdot7/2},$$

$$L_{11s3}=\sinh\left[\frac{x}{2}: \right.\frac{4\cdot5
\dots(k+n+3)}
{3\cdot 4\dots\ (2+k+n)}=\sinh\left[\frac{x}{2}: \right.\frac{3+k+n}
{3},$$

$$L_{01}=\sinh\left[\frac{x}{2}: \right. 
\frac{1+n}{1}$$

$$L_{c2}=1$$

The product of all these terms

     \begin{multline*}
    X(x,k,n,-2,5,6) = \\
         \sinh\left[\frac{x}{2}: \right.
         \frac{(n+1)(3+k+n)^2(5/2+k+n)(7/2+k+n)(2+k+n)(4+k+n)(2+k)^2
    }
     {1^2\cdot2^3\cdot3^3\cdot 4^2\cdot 5^2}\times\\
    \times \frac{ (2k+8)(4k+8)(4k+10)^2(k+1)(k+3)(4+2k+n)(5+2k+n)^2(6+2k+n)}{6\cdot 7\cdot8\cdot 3/2 \cdot(5/2)^2}\times\\
   \times \frac{(7+3k+n)(8+3k+2n)(3/2+k)(5/2+k)(9/2+2k+n)(11/2+2k+n)}{7/2\cdot9/2\cdot11/2}
     \end{multline*}    
This immediately coincides with the expression Weyl formula gives for $\lambda=k\omega_2+n\omega_1$ highest weight representations of $F_4$ 
 algebra.
 
 \subsection{$E_6$}
 For $E_6$ the Vogel parameters are $\alpha=-2, \beta=6, \gamma=8$.

$$L_{31}=\sinh\left[\frac{x}{2}: \right.
\frac{3k+n+10}{10},$$

$$L_{32}=\sinh\left[\frac{x}{2}: \right.\frac{3k+2n+11}{11}
,$$

$$L_{21s1}=\sinh\left[\frac{x}{2}: \right.\frac{10\cdot11
\dots(9+2k+n)}
{5\cdot 6\dots\ (4+2k+n)}=\sinh\left[\frac{x}{2}: \right.\frac{(5+2k+n)\dots(9+2k+n)}
{5\cdot6\dots9},$$

$$L_{21s2}=\sinh\left[\frac{x}{2}: \right.\frac{9\cdot10
\dots(8+2k+n)}
{6\cdot 7\dots\ (5+2k+n)}=\sinh\left[\frac{x}{2}: \right.\frac{(6+2k+n)(7+2k+n)(9+2k+n)}
{6\cdot7\cdot8}
,$$

$$L_{21s3}=\sinh\left[\frac{x}{2}: \right.\frac{7+2k+n}{7}
,$$

$$L_{10s1}=\sinh\left[\frac{x}{2}: \right.\frac{6\cdot7
\dots(5+k)}
{1\cdot 2\dots\ k}=\sinh\left[\frac{x}{2}: \right.\frac{(1+k)\dots(5+k)}
{1\cdot2\dots5},$$

$$L_{10s2}=\sinh\left[\frac{x}{2}: \right.\frac{5\cdot6
\dots(4+k)}
{2\cdot 3\dots\ (1+k)}=\sinh\left[\frac{x}{2}: \right.\frac{(2+k)(3+k)(4+k)}
{2\cdot3\cdot4},$$

$$L_{10s3}=\sinh\left[\frac{x}{2}: \right.\frac{4\cdot5
\dots(k+3)}
{3\cdot 4\dots\ (k+2)}=\sinh\left[\frac{x}{2}: \right.\frac{k+3}
{3},$$

$$L_{11s1}=\sinh\left[\frac{x}{2}: \right.\frac{7\cdot8
\dots(6+k+n)}
{2\cdot 3\dots\ (k+n+1)}=\sinh\left[\frac{x}{2}: \right.\frac{(k+n+2)\dots(k+n+6)}
{2\cdot3\dots6},$$

$$L_{11s2}=\sinh\left[\frac{x}{2}: \right.\frac{6\cdot7
\dots(5+k+n)}
{3\cdot 4\dots\ (2+k+n)}=\sinh\left[\frac{x}{2}: \right.\frac{(3+k+n)(4+k+n)(5+k+n)}
{3\cdot4\cdot5},$$

$$L_{11s3}=\sinh\left[\frac{x}{2}: \right.\frac{4+k+n}
{4},$$

$$L_{01}=\sinh\left[\frac{x}{2}: \right. 
\frac{1+n}{1}$$

$$L_{c2}=1$$
The product of all these terms gives
  \begin{multline*}
    X(x,k,n,-2,6,8) = \\
         \sinh\left[\frac{x}{2}: \right.
         \frac{(k+1)(k+2)^2(k+3)^2(k+4)^2(k+5)(n+1)(2+k+n)
    }
     {1^2\cdot2^3\cdot3^4\cdot 4^5}\times\\
    \times \frac{ (3+k+n)^2(4+k+n)^3(5+k+n)^2(6+k+n)(5+2k+n)(6+2k+n)^2(7+2k+n)^3}{5^4\cdot 6^3\cdot7^3\cdot 8^2}\times\\
   \times \frac{(8+2k+n)^2(9+2k+n)(10+3k+n)(11+3k+2n)}{9\cdot10\cdot11}
     \end{multline*}

     which coincides with the quantum dimension (\ref{W}) of the $\lambda=k\omega_3+n\omega_6$ irrep .
     
     \subsection{$E_7$}
 For $E_7$ Vogel's parameters are $\alpha=-2, \beta=8, \gamma=12$.
$$L_{31}=\sinh\left[\frac{x}{2}: \right.
\frac{3k+n+16}{16},$$

$$L_{32}=\sinh\left[\frac{x}{2}: \right.\frac{3k+2n+17}{17}
,$$

$$L_{21s1}=\sinh\left[\frac{x}{2}: \right.\frac{16\cdot17
\dots(15+2k+n)}
{7\cdot 8\dots\ (6+2k+n)}=\sinh\left[\frac{x}{2}: \right.\frac{(7+2k+n)\dots(15+2k+n)}
{7\cdot8\dots15},$$

$$L_{21s2}=\sinh\left[\frac{x}{2}: \right.\frac{14\cdot15
\dots(13+2k+n)}
{9\cdot 10\dots\ (8+2k+n)}=\sinh\left[\frac{x}{2}: \right.\frac{(9+2k+n)\dots(13+2k+n)}
{9\cdot10\dots13}
,$$

$$L_{21s3}=\sinh\left[\frac{x}{2}: \right.\frac{11+2k+n}{11}
,$$

$$L_{10s1}=\sinh\left[\frac{x}{2}: \right.\frac{10\cdot11
\dots(9+k)}
{1\cdot 2\dots\ k}=\sinh\left[\frac{x}{2}: \right.\frac{(1+k)\dots(9+k)}
{1\cdot2\dots9},$$

$$L_{10s2}=\sinh\left[\frac{x}{2}: \right.\frac{8\cdot9
\dots(7+k)}
{3\cdot 4\dots\ (2+k)}=\sinh\left[\frac{x}{2}: \right.\frac{(3+k)\dots(7+k)}
{3\cdot4\dots7},$$

$$L_{10s3}=\sinh\left[\frac{x}{2}: \right.\frac{6\cdot7
\dots(k+5)}
{5\cdot 6\dots\ (k+4)}=\sinh\left[\frac{x}{2}: \right.\frac{k+5}
{5},$$

$$L_{11s1}=\sinh\left[\frac{x}{2}: \right.\frac{11\cdot12
\dots(10+k+n)}
{2\cdot 3\dots\ (k+n+1)}=\sinh\left[\frac{x}{2}: \right.\frac{(k+n+2)\dots(k+n+10)}
{2\cdot3\dots10},$$

$$L_{11s2}=\sinh\left[\frac{x}{2}: \right.\frac{9\cdot10
\dots(8+k+n)}
{4\cdot 5\dots\ (3+k+n)}=\sinh\left[\frac{x}{2}: \right.\frac{(4+k+n)\dots(8+k+n)}
{4\dots8},$$

$$L_{11s3}=\sinh\left[\frac{x}{2}: \right.\frac{6+k+n}
{6},$$

$$L_{01}=\sinh\left[\frac{x}{2}: \right. 
\frac{1+n}{1}$$

$$L_{c2}=1$$
The product of all these terms gives
  \begin{multline*}
    X(x,k,n,-2,8,12) = \\
         \sinh\left[\frac{x}{2}: \right.
         \frac{(k+1)(k+2)(k+3)^2\dots(k+7)^2(k+8)(k+9)(7+2k+n)
    }
     {1^2\cdot2^3\cdot3^3\cdot 4^2\cdot 5^5}\times\\
     \frac{(8+2k+n)(9+2k+n)^2(10+2k+n)^2(11+2k+n)^3(12+2k+n)^2(13+2k+n)^2(14+2k+n)}{ 6^5\cdot7^5\cdot8^4}\\
    \times \frac{ (15+2k+n)(16+3k+n)(17+3k+2n)(1+n)(5+k)(2+k+n)(3+k+n)(4+k+n)^2}{ 9^4\cdot10^3\cdot 11^3\cdot12^2}\times\\
   \times \frac{(5+k+n)^2(6+k+n)^3(7+k+n)^2(8+k+n)^2(9+k+n)(10+k+n)}{13^2\cdot14\cdot15\cdot16\cdot17}
     \end{multline*}

     which coincides with quantum dimension of the $\lambda=k\omega_2+n\omega_1$ irrep.
     
     \subsection{$E_8$}
 For $E_8$ the Vogel parameters are $\alpha=-2, \beta=12, \gamma=20$.
$$L_{31}=\sinh\left[\frac{x}{2}: \right.
\frac{3k+n+28}{28},$$

$$L_{32}=\sinh\left[\frac{x}{2}: \right.\frac{3k+2n+29}{29}
,$$

$$L_{21s1}=\sinh\left[\frac{x}{2}: \right.\frac{28\cdot29
\dots(27+2k+n)}
{11\cdot 12\dots\ (10+2k+n)}=\sinh\left[\frac{x}{2}: \right.\frac{(11+2k+n)\dots(27+2k+n)}
{11\cdot12\dots27},$$

$$L_{21s2}=\sinh\left[\frac{x}{2}: \right.\frac{24\cdot25
\dots(23+2k+n)}
{15\cdot 16\dots\ (14+2k+n)}=\sinh\left[\frac{x}{2}: \right.\frac{(15+2k+n)\dots(23+2k+n)}
{15\cdot16\dots23}
,$$

$$L_{21s3}=\sinh\left[\frac{x}{2}: \right.\frac{19+2k+n}{19}
,$$

$$L_{10s1}=\sinh\left[\frac{x}{2}: \right.\frac{18\cdot19
\dots(17+k)}
{1\cdot 2\dots\ k}=\sinh\left[\frac{x}{2}: \right.\frac{(1+k)\dots(17+k)}
{1\cdot2\dots17},$$

$$L_{10s2}=\sinh\left[\frac{x}{2}: \right.\frac{14\cdot15
\dots(13+k)}
{5\cdot 6\dots\ (4+k)}=\sinh\left[\frac{x}{2}: \right.\frac{(5+k)\dots(13+k)}
{5\cdot6\dots13},$$

$$L_{10s3}=\sinh\left[\frac{x}{2}: \right.\frac{10\cdot11
\dots(k+9)}
{9\cdot 10\dots\ (k+8)}=\sinh\left[\frac{x}{2}: \right.\frac{k+9}
{9},$$

$$L_{11s1}=\sinh\left[\frac{x}{2}: \right.\frac{19\cdot20
\dots(18+k+n)}
{2\cdot 3\dots\ (k+n+1)}=\sinh\left[\frac{x}{2}: \right.\frac{(k+n+2)\dots(k+n+18)}
{2\cdot3\dots18},$$

$$L_{11s2}=\sinh\left[\frac{x}{2}: \right.\frac{15\cdot16
\dots(14+k+n)}
{6\cdot 7\dots\ (5+k+n)}=\sinh\left[\frac{x}{2}: \right.\frac{(6+k+n)\dots(14+k+n)}
{6\cdot7\dots14},$$

$$L_{11s3}=\sinh\left[\frac{x}{2}: \right.\frac{10+k+n}
{10},$$

$$L_{01}=\sinh\left[\frac{x}{2}: \right. 
\frac{1+n}{1}$$

$$L_{c2}=1$$
     
       \begin{multline*}
    X(x,k,n,-2,8,12) = \\
         \sinh\left[\frac{x}{2}: \right.
         \frac{(n+1)(2+k+n)(3+k+n)\dots(6+k+n)^2(7+k+n)^2\dots(10+k+n)^3
    }
     {1^2\cdot2^3\cdot3^2\cdot 4^2\cdot 5^3\cdot6^4\cdot7^4\cdot8^4\cdot9^5\cdot10^5}\times\\
     \frac{(11+k+n)^2\dots(14+k+n)^2(15+k+n)\dots(18+k+n)^2(1+k)(2+k)\dots(5+k)^2(6+k)^2\dots(9+k)^3}{ 11^5\cdot12^5\cdot13^5\cdot14^4\cdot15^4\cdot16^4\cdot17^4\cdot18^3\cdot19^3\cdot20^2}\\
    \times \frac{ (10+k)^2\dots(13+k)^2(14+k)\dots(17+k)(11+2k+n)\dots(15+2k+n)^2\dots(19+2k+n)^3}{ 9^4\cdot10^3\cdot 11^3\cdot12^2}\times\\
   \times \frac{(20+2k+n)^2\dots(23+2k+n)^2(24+2k+n)\dots(27+2k+n)(28+3k+n)(29+3k+2n)}{21^2\cdot22^2\cdot23^2\cdot24\cdot25\cdot26\cdot27\cdot28\cdot29}
     \end{multline*}    
      coinciding with direct calculation by (\ref{W}) carried out for $\lambda=k\omega_6+n\omega_7$ irrep.


\begin{thebibliography}{99}
	
	\bibitem{V0}
	P. Vogel,  The Universal Lie algebra. Preprint (1999), https://webusers.imj-prg.fr/\~{}pierre.vogel/grenoble-99b.pdf
	
	
	\bibitem{M16QD}
	R.L.Mkrtchyan, On Universal Quantum Dimensions, arxiv:1610.09910, Nuclear Physics B921,  2017, pp. 236-249, 
	
	\bibitem{LM1} 
	J.M. Landsberg and  L.Manivel,  A universal dimension formula for complex simple Lie algebras. Adv. Math. {\bf 201} (2006), 379-407
	
	
	\bibitem{King}
        R. C. King {\it The dimensions of irreducible tensor representations of the orthogonal and symplectic groups.} Can. J. Math. {\bf 33} (1972), 176.


         \bibitem{MV11} 
	Ruben L. Mkrtchyan and Alexander P. Veselov, On duality and negative dimensions in the theory of Lie groups and symmetric spaces, arxiv:1011.0151, Journal Math. 
	Phys., 52, 083514 (2011)
	
	
	\bibitem{Cvit} 
	P. Cvitanovic and A. D. Kennedy {\it Spinors in negative dimensions.} Phys. Scr. {\bf 26} (1982), 5-12.
	
	
	\bibitem{Cohen}
	A. M.Cohen and  R. de Man,   Computational evidence for Deligne's conjecture regarding exceptional Lie
	groups, Comptes Rendus de l'Académie des Sciences, Série 1, Mathématique, (1996) 322(5), 427-432
	
	
	\bibitem{FK}
        Robert Fegera and Thomas W. Kephart, LieART – A Mathematica Application for Lie Algebras and Representation Theory, arxiv:1206.6379
        
        
        \bibitem{DiF} 
	P. Di Francesco, P.Mathieu   and  D.S\'en\'echal,  Conformal Field Theory. Springer-Verlag, New York, 1997. 
	
	
	 \bibitem{AM}
	M.Y. Avetisyan and R.L. Mkrtchyan,  $X_2$ Series of Universal Quantum Dimensions, arXiv:1812.07914
	
	
	\bibitem{Cvitbook}
	P. Cvitanovic {\it Group Theory.} Princeton University Press, Princeton, NJ, 2004. 
	http://www.nbi.dk/group theory
	
	
	\bibitem{Mkr}
	R.L. Mkrtchyan {\it The equivalence of $Sp(2N)$ and $SO(-2N)$ gauge theories.} Physics Letters {\bf 105B} (1981), 174-176.
	
	
	\bibitem{MSV}
	R.L. Mkrtchyan, A.N. Sergeev and A.P. Veselov,  
	Casimir eigenvalues for universal Lie algebra, arxiv:1105.0115,  Journ. Math.Phys. 53, 102106 (2012).
	
	
	\bibitem{A19}	
        M.Y. Avetisyan, On Universal Eigenvalues of Casimir Operator, arXiv:1908.08794.
        
        
        \bibitem{V}
	P.Vogel,   Algebraic structures on modules of diagrams. Preprint (1995),  www.math.jussieu.fr/\~{}vogel/diagrams.pdf, J. Pure Appl. Algebra {\bf 215} (2011), no. 6, 1292-1339. 
	
	
	\bibitem{M13}
	R.L.Mkrtchyan,  Nonperturbative universal Chern-Simons theory.  JHEP09(2013)54, arxiv:1302.1507.
	
	
	\bibitem{KM}
	D. Krefl and  R.Mkrtchyan, Exact Chern-Simons / Topological String duality, arxiv:1506.03907, JHEP10, (2015), 45.
	
	\bibitem{MV}
	R.L. Mkrtchyan  and A.P.Veselov,  Universality in Chern-Simons theory. JHEP08 (2012) 153, arxiv:1203.0766.

	\bibitem{MMM}
	A. Mironov, R. Mkrtchyan and A. Morozov,  On universal knot polynomials, arxiv:1510.05884 , JHEP02(2016)078
	
	
	
	
	
	

       	
	

\end{thebibliography}
\end{document}